\documentclass{elsart}
\journal{Physics Reports}
\usepackage{epsfig}

\def\be{\begin{eqnarray}}
\def\ee{\end{eqnarray}}
\def\lsim{\mathrel{\rlap{\lower3pt\hbox{\hskip1pt$\sim$}}
     \raise1pt\hbox{$<$}}} 
\def\gsim{\mathrel{\rlap{\lower3pt\hbox{\hskip1pt$\sim$}}
     \raise1pt\hbox{$>$}}} 
\def\la{\langle}\def\ra{\rangle}

\def\cal{\it}

\begin{document}

\runauthor{Brown, Lee \& Rho}

\begin{frontmatter}
\title{Nature of the Chiral Restoration Transition \\ in QCD}

\author[suny]{Gerald E. Brown,\thanksref{geb}}
\author[suny,ipnl]{Lo\"{\i}c Grandchamp,\thanksref{loic}}
\author[pnu]{Chang-Hwan Lee,\thanksref{chl}}
\author[saclay,kias]{Mannque Rho\thanksref{rho}}

\address[suny]{Department of Physics and Astronomy,
               State University of New York, Stony Brook, NY 11794, USA}
\address[ipnl]{IPN Lyon, IN2P3-CNRS et UCBL, 43 Bd. du 11 Novembre
  1918, 69622 Villeurbanne Cedex, France}
\address[pnu]{Department of Physics, Pusan National University,
              Pusan 609-735, Korea}
\address[saclay]{Service de Physique Th\'eorique, CEA/DSM/SPhT. Unit\'e de
recherche associ\'ee au CNRS, CEA/Saclay, 91191 Gif-sur-Yvette c\'edex, France}
\address[kias]{School of Physics, Korea Institute for Advanced Study,
               Seoul 130-722, Korea}
\thanks[geb]{Ellen.Popenoe@sunysb.edu}
\thanks[loic]{loic@tonic.physics.sunysb.edu}
\thanks[chl]{clee@pusan.ac.kr}
\thanks[rho]{rho@spht.saclay.cea.fr}

\begin{abstract}
As the chirally restored phase ends with $T$ coming down to $T_c$, a
phase resembling a mixed phase is realized, during which the hadrons
(which are massless at $T_c$ in the chiral limit) get their masses
back out of their kinetic energy. The gluon condensation energy is fed
into the system to keep the temperature (nearly) constant. Lattice
results for the gluon condensation are matched by a Nambu-Jona-Lasinio
calculation. The latter shows that below $T_c$ the chiral
symmetry is barely broken, so that with an $\sim 6\%$ drop in the
scalar coupling $G$ it is restored at $T_c$. Nearly half of the glue,
which we call epoxy, is not melted at $T_c$.
\end{abstract}
\begin{keyword}
\PACS{97.60.Lf; 97.80.Jp}
\end{keyword}

\end{frontmatter}

\section{Intoduction\label{intro}}

As the chirally restored phase of the QGP plasma ends, it is
commonly assumed that there is a mixed phase of that plasma with
hadronic excitations with constant temperature before freezeout of
the hadrons. We will show, however, that in the chiral limit as
one approaches $T_c$, the plasma consists of {\em nearly} massless
hadrons. As these go back on mass shell, chiefly using their
kinetic energy to do so, the temperature drops only slightly - at
least the freeze-out temperature is only a few MeV below $T_c$. In
fact, the system is assumed to be equilibrated at $T_c$, and also
at freezeout, where the interactions are strong enough to
equilibrate it. It may well go free-flow between $T_c$ and
$T_{freeze\; out}$, the condensation energy of soft glue
furnishing the ``dark energy" to produce the scalar field energy
in the hadronic phase.

\section{Nature of the Scale Anomaly}

Harada \& Yamawaki \cite{Harada03} have shown in putting into
renormalization group (RG) equations the hidden local symmetry
theory matched at a suitable scale to QCD that in the chiral limit
hot/dense matter flows to the ``vector manifestation (VM) fixed
point" at $m_\rho^\star=0$ and $f_\pi^\star=0$ as chiral symmetry
is restored. This provides a support for ``Brown-Rho (BR)
scaling"~\cite{BRscaling} which claims that light-quark hadron
masses go to zero with chiral restoration~\footnote{\small It is
perhaps worthwhile to give more precision in light of the more
recent development to the notion of BR scaling. The original
formulation~\cite{BRscaling} was based on the skyrmion Lagrangian
implemented with the scale anomaly of QCD embedded in hadronic
medium. The scaling behavior obtained there was at a mean field
level and at that level, the ``parametric mass" of the Lagrangian
and the ``pole mass" inside the medium, both of which are
dependent on the background (temperature and/or density), were the
same. As recently explained~\cite{BRrecent,BR02}, the important
development by Harada and Yamawaki shows however that what
represents BR scaling is the {\it parametric dependence} in the
bare Lagrangian which governs the background dependence of the
parametric masses (and coupling constants) when quantum effects
and thermal or dense loop corrections are taken into account. At
the vector manifestation (VM) fixed point, the pole mass of the
vector meson does vanish because the parametric mass and coupling
constant vanish at that point. This modern interpretation of BR
scaling is reformulated in terms of the skyrmion description of
dense matter in \cite{LPRV}.

Another important point that we should keep in mind is that it
does not make sense to consider a system ``sitting" right on top
of the VM point~\cite{Harada03}: It makes sense only to {\it
approach} it from the broken symmetry sector. It is in this sense
that we shall refer to ``massless hadrons" in what follows.}.
Shuryak \& Brown \cite{Shuryak03} have recently discussed evidence
for the decrease {\it in medium} of the $\rho$-meson mass by the
STAR experiments \cite{STARexp}.

In this note we wish to understand what happens to the scale
anomaly of QCD with restoration of chiral symmetry. As laid out 
first by Adami, Hatsuda \& Zahed\cite{adami} and later by
Koch \& Brown \cite{Koch93}, this anomaly consists of
about 50/50 of soft glue, which condenses as the temperature of
the quark-gluon plasma moves downwards through $T_c$ and hard
glue, or ``epoxy", which condenses slowly over a wide range of
temperatures, and has little effect on the phase transition. The
epoxy can be described as residing in instanton molecules, as we
discuss below.

We use the Gross-Neveu model in four dimensions, that is,
essentially Nambu-Jona-Lasinio without pion, as in Brown, Buballa
\& Rho \cite{Brown96} that we shall refer to as BBR. The
Lagrangian is
 \be {\cal L} =\bar\psi
(i\partial\!\!\!/+g\sigma)\psi-\frac 12 m_\sigma^2 \sigma^2.
\label{eq1}
 \ee Here $\sigma$ is an auxiliary (scalar) field. Its
vacuum expectation value can be obtained by setting the variation
of the expectation value of ${\cal L}$ to zero
 \be \delta\la {\cal
L}\ra/\delta\sigma=0 \ee from which we obtain \be
\sigma=\frac{g}{m_\sigma^2}\langle\bar\psi\psi\rangle. \label{eq3}
 \ee
Eq.~(\ref{eq1}) with eq.(\ref{eq3}) looks very much like Walecka
theory at mean-field level, except that negative-energy states are
included in the $\langle\bar\psi\psi\rangle$ of Eq.~(\ref{eq3})
and one must take their kinetic energy into account.

As noted in BBR, the proper variables for $\rho=0$, $T=0$ are
nucleons. However, the lowering of the energy of the negative
energy sea because of the mass generation (cut off at $\Lambda$)
is \cite{Brown88} \be
{B.E.}(glue)=-E_{vac}=4\int_0^\Lambda\frac{d^3k}{(2\pi)^3}\sqrt{k^2+m_N^2}
-\frac{\Lambda^4}{2\pi^2}.
 \ee
This equation differs from Eq.~(4.10) of BBR, which had an
additional term $-\frac 12 m_\sigma\sigma^2$, which subtracted off
the field energy of the $T=0$, $\rho=0$ configuration. This $\frac
12 m_\sigma^2\sigma^2$ will be fed back into the system as
$\sigma$ goes to zero, but will chiefly heat up the pions and
other particles, and will not have any appreciable effect on the
constituent quarks.

With $\Lambda=660$ MeV and the NJL $G\Lambda^2=4.3$, which give
$T_c=170$ MeV, we find $B.E.(glue)=0.012\ {\rm GeV}^4$. This is
the correct total gluon condensate. However, it does not agree
with the Miller lattice calculation, which we discuss in the next
section. The reason for this may be understood in terms of the
random instanton vacuum structure~\cite{schaefer95}. In the random
instanton vacuum picture, there is considerable rearrangement of
the glue in the vacuum at low temperatures into instanton
molecules by $T_c$. These molecules with nearest neighbor
instanton just fill the compacted time dimension of $\pi/T$ for
$T_c$. The $\sim 50\%$ glue in the molecules does not melt with
chiral restoration, making up the epoxy background.
As discussed by BBR, at some point in temperature or density,
nucleons dissociate smoothly into constituent quarks, with
$m_Q^\star\sim m_N^\star/3$. When they are fully dissociated, we
will have the NJL correlation energy ( which we identify with bag
constant)
 \be {B.E.}(soft\;
glue)=12\int_0^\Lambda\frac{d^3k}{(2\pi)^3}\sqrt{k^2+m_Q^2}
-\frac{3\Lambda^4}{2\pi^2} \label{eq5}
 \ee
the $\Lambda$ being the same. The $\Lambda$ will be chosen so as
to give $T_c\sim 170$ MeV under the conditions for RHIC
experiments.

Now we assume that appreciable melting of the soft glue occurs at
temperatures high enough that the nucleons have dissociated into
constituent quarks. We find in the case of lattice calculations of
charmonium\cite{Karsch01} that their curve bends flat at $T_c$,
with a drop of 640 MeV, indicating that the constituent quark mass
of the nonstrange quark is $\sim 320$ MeV. (See also
Ref.~\cite{Grandchamp} for this interpretation.) With
$\Lambda=660$ MeV and $m_Q=320$ MeV we find from Eq.~(\ref{eq5})
that \be B.E(soft\; glue)=0.0058\ {\rm GeV}^4,
 \ee
roughly 50\% of the total glue.

Now the only way that NJL knows about the gluon interactions is
through $G$ as the glue degrees of freedom have been integrated
out~\footnote{We thank M. Prakash for stressing this point.}. As
the soft glue is melted, $G$ should decrease from the value at which it can break
chiral symmetry. However, at that point a new NJL-type
interaction, chiefly driven by the instanton-anti-instanton
molecules arises \cite{schaefer95}. Since about half of the
original glue, the epoxy, is retained in those molecules, one
would not expect the original $G$, which below $T_c$ broke chiral
symmetry enough to condense the other half of the glue, to be much
larger than the $G$ slightly above $T_c$.

\section{Lattice Results}

Calculations including both light dynamical quarks and heavier
ones have been carried out by D.E. Miller \cite{Miller00} and show
that the trace anomaly decreases from $0.012 $ GeV$^4$ at $T\sim
122$ MeV to $0.0056\ {\rm GeV}^4$ at $T\sim 170$ MeV. We take the
latter to be $T_c$ for his unquenched calculations. In other
words, the drop in total bag constant at $T_c$ is 53\% of the
total gluon condensate, the remainder being ``hard glue" or
``epoxy".

\begin{figure}
\centerline{\epsfig{file=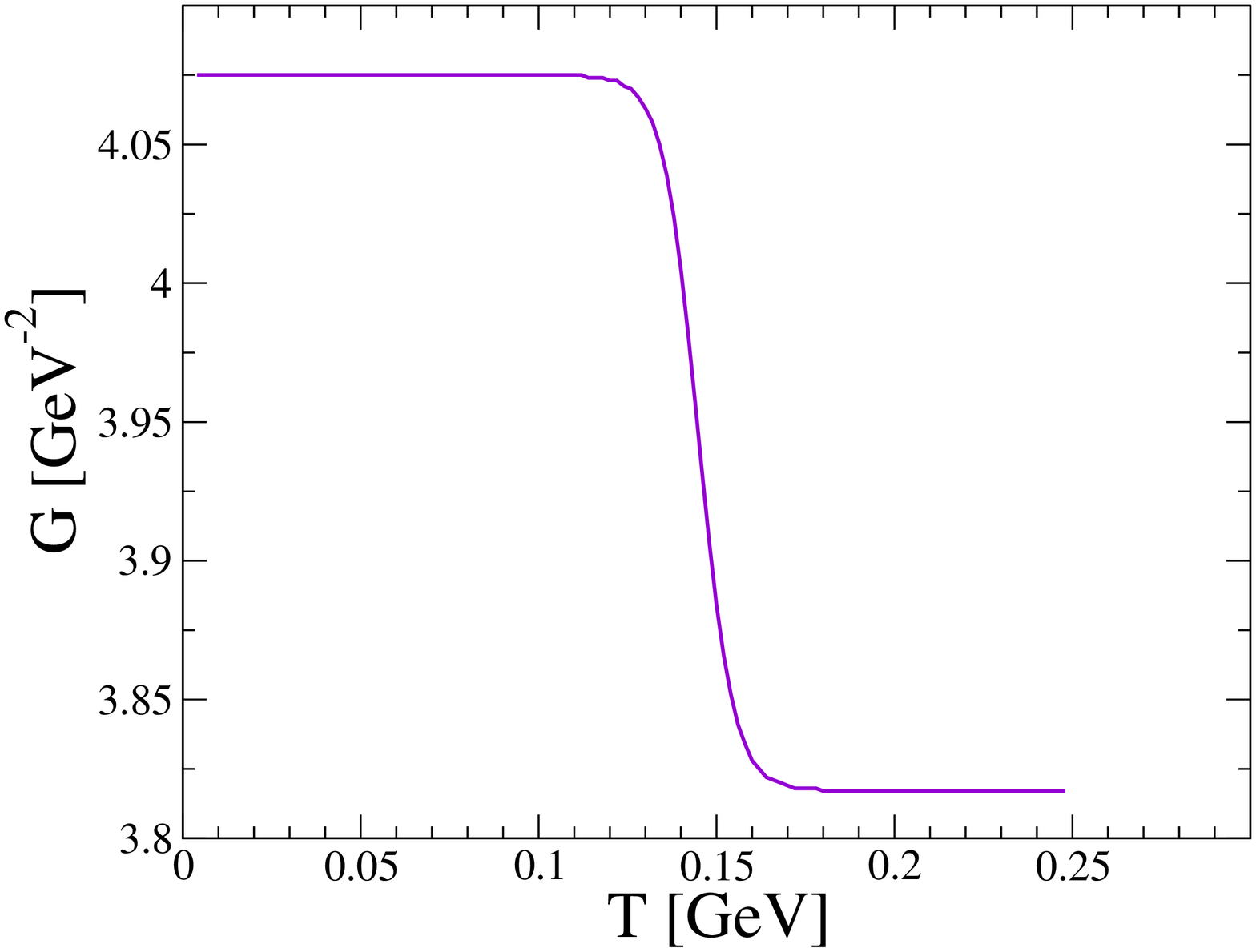,height=3in}}
\centerline{\epsfig{file=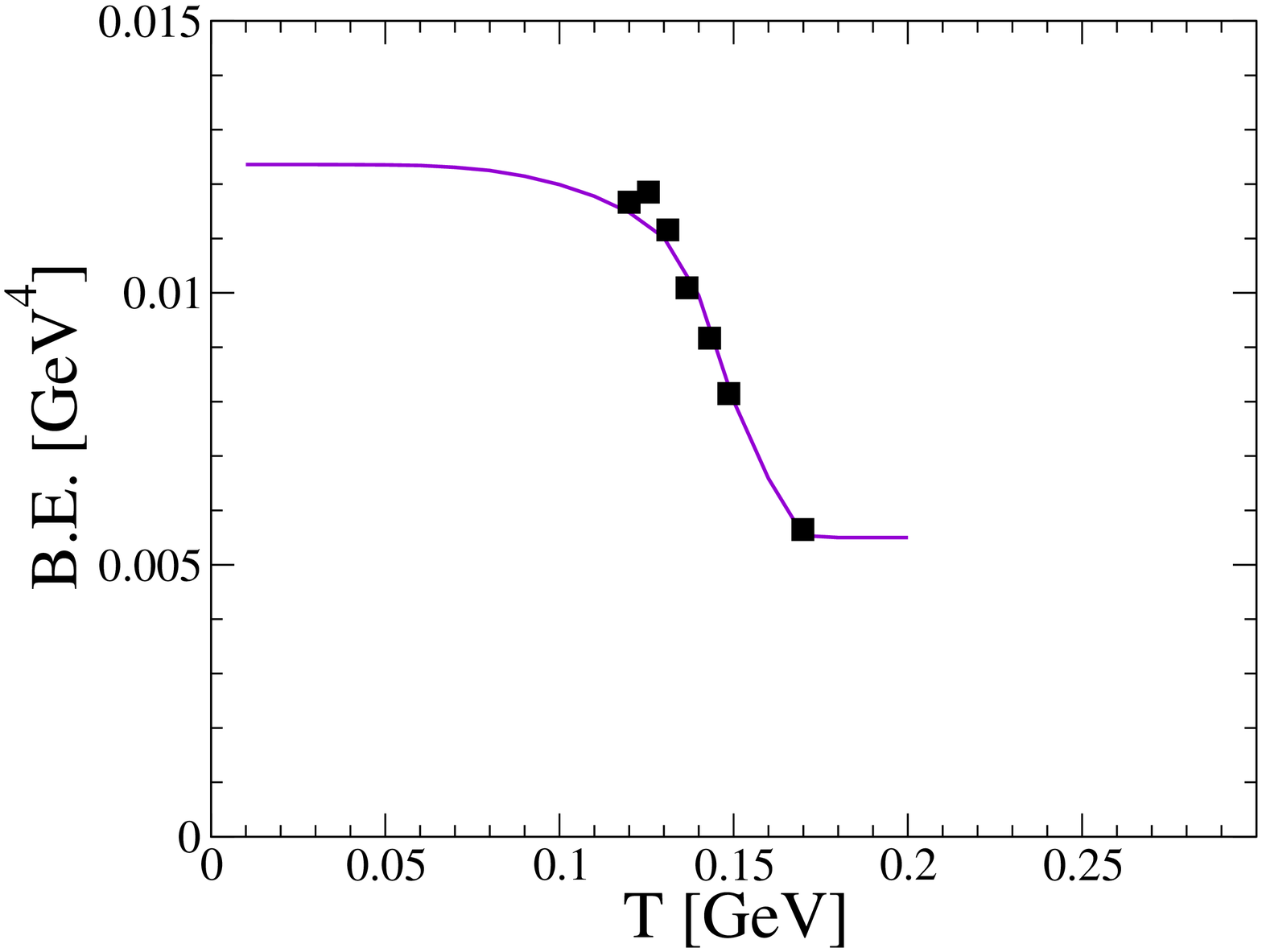,height=3in}} \caption{(a)
Temperature dependence of the coupling $G$ and (b) $B.E.(soft\;
glue)$ with varying $G$. In this calculation $\Lambda=695$ MeV and
$m_{Q,current}=0$ are used.} \label{fig3}
\end{figure}

\begin{figure}
\centerline{\epsfig{file=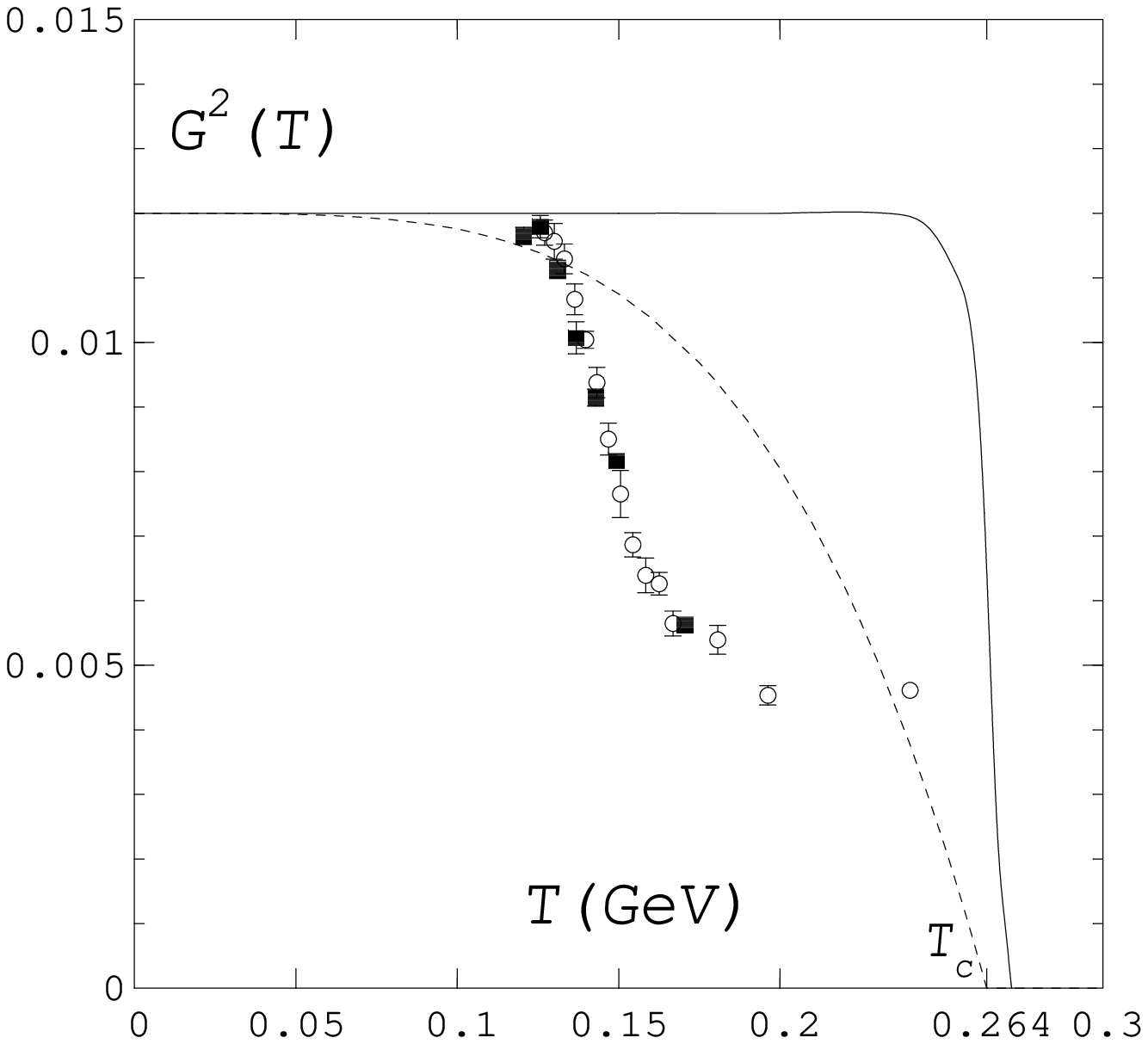,height=3in,
bbllx=85,bblly=240,bburx=484,bbury=611}}
\caption{Gluon
condensates taken from Miller\cite{Miller00}. The lines show the
trace anomaly for SU(3) (solid) and the ideal gluon gas (broken)
in comparison with that of the light dynamical quarks denoted by
the open circles and the heavier ones by filled circles. The $T_c$
marked in the figure is that for quenched QCD, whereas we deal
with unquenched QCD in this note.} \label{fig1}
\end{figure}

In Fig.~\ref{fig3} we see that an $\sim 6\%$ drop in $G$
in the neighborhood of $T_c$ allows us to fit Miller's lattice results of
Fig.~\ref{fig1} quite well.
Although it may be a coincidence, this $\sim 6\%$ is roughly how much
greater the soft glue is than the epoxy in Miller's results.

Our scenario is, then:
\begin{enumerate}

\item At $T\sim 125$ MeV, the nucleons have begun to ``loosen" into
constituent quarks. About half of the binding energy of the negative
energy nucleons goes into the formation of instanton molecules
(46\% in Miller's lattice calculation which are shown below).
The remaining binding energy which is the negative energy of
constituent quarks is melted as soft glue.

\item As $T$ increases above 150 MeV, the coupling constant $G$ which contains the
information about the gluonic interactions drops, so that by
$T_{f.o.}\sim 170$ MeV it is below the value needed to break
chiral symmetry. Only an $\sim 6\%$ drop is needed for the
interaction to no longer break chiral symmetry. Thus, the new
Nambu-Jona-Lasinio~\cite{schaefer95} formulated about the
instanton molecules which no longer break chiral symmetry would be
expected to have coupling constants for the scalar, pseudoscalar,
vector and axial-vector excitations which are just about as large
as those of the original NJL in the symmetry-broken sector. The
fact that such a small change in $G$ is needed in going from the
chirally broken to restored phase is compatible with the $\sim
50/50$ division of glue into soft and hard.

\end{enumerate}

The hard glue is melted at temperatures $T>T_c$ over a much greater
scale, and does not concern our arguments for $T<T_c$ here.

\section{Chiral Symmetry Breaking}

As noted earlier, Harada \& Yamawaki\cite{Harada03} have shown
that hadron masses go to zero in the chiral limit. Brown \& Rho
\cite{BR02} showed that the scaling of hadron masses could be
realized in terms of the scaling of the NJL $\langle\bar q
q\rangle^\star$, the star denoting the {\it in-medium} condensate.

The $G^\prime\equiv G\Lambda^2$ in NJL is, within
factors,\footnote{In fact, from the particle data tables
$m_\sigma=m_{f(0)}=600$ MeV is not far from our $\Lambda=660$
MeV.} $(g_{\sigma QQ}/m_\sigma^2)\Lambda^2$ in the constituent
quark sector. As suggested by Brown \& Rho\cite{BR02} the
$\Lambda$ may be identified as the Wilsonian matching scale for
constituent quarks, and should therefore be taken to be
independent of density or temperature. However $m_\sigma^\star$
should scale with the masses of the other mesons. Thus, to the
extent that ${g^\star_{\sigma QQ}}/{m_\sigma^\star}$ does not
change, we see that $g_{\sigma QQ}^\star$ scales roughly as
${m_Q^\star}$, the ratio dropping at most only a few percent with
chiral restoration. The NJL in the instanton molecule sector for
$T>T_c$~\cite{schaefer95} has, therefore, nearly the same coupling
constants as the NJL in the broken symmetry sector, although those
in the former are not quite strong enough to break chiral
symmetry. In fact, the total amount of glue, 0.12 GeV$^4$, assumed
by Miller~\cite{Miller00} has been superseded by somewhat larger
values~\cite{largerglue}. However Miller's determination of the
soft glue which is melted by $T_c$ remains valid. What is changed
is the amount of hard glue left unmelted and left possibly in the
form of epoxy.

In the instanton description~\cite{schaefer95}, the hard glue is
in the form of instanton molecules.  Our development above
suggests that at chiral restoration, which brings the system into
a ``liquid" of instanton molecules for $T\gsim T_c$, the
interactions change little. In the instanton-molecule NJL, the
quark masses from the breaking of chiral symmetry below $T_c$ are
replaced by thermal masses (given in perturbation theory by
$m_Q=gT/\sqrt{6}$~\cite{X}) above $T_c$.

In a recent publication, Shuryak and Zahed~\cite{shuryakzahed}
have suggested that the evolution of the color Coulomb interaction
is such that light quark states are likely to be Coulomb-bound up
to a temperature of $T_{\bar{q}q}\approx 1.45\ T_c\approx 250$
MeV. This color Coulomb interaction is not incorporated in the
zero-ranged instanton-molecule NJL four-Fermi interactions and
should therefore be added in the theory. The
attractive four-Fermi interactions strengthen the binding such as
to support light-quark bound states up to an even higher
temperature. We are currently in the process of implementing the
color Coulomb interaction to the NJL Lagrangian, but the general
result is already clear; i.e., RHIC has found the Instanton
Molecule Liquid, not the quark-gluon plasma.

The $q\bar{q}$ and $gg$ bound states for $T\gsim T_c$ will have
the very important effects that Shuryak and Zahed have emphasized.
As the fire-ball formed in RHIC collisions expands and the initial
temperature decreases, these bound states will begin to form. Just
as they go through zero binding energy, their scattering amplitude
$a$ will go to infinity, and their scattering cross section
$\sigma=4\pi a^2$, also. The net result of these large cross
sections will be a low viscosity and good equilibration. Instead
of a quark-gluon plasma, we will have ``sticky molasses."
Note that the Shuryak \& Zahed $\bar q q$ and $gg$ bound states
are colorless, so that color is filtered out in their formation;
i.e., the Instanton Molecule Liquid is colorless.

Now going from $T_c$ downwards in temperature, the phase in which
the hadrons get their masses back has many of the properties of a
mixed phase, but it is not a genuine mixed phase, since it is
composed of (``off-shell"~\footnote{Note that the hadrons are
off-shell with respect to the low-temperature environment in which
the masses are measured. The hadrons are actually ``on-shell" in
the heat bath, their masses being the pole masses in the thermal
Green's functions.}) hadrons. Nonetheless, the temperature changes
only little until freezeout, although the hadrons get most of
their mass back, because ``dark energy" is fed back through the
dropping field energy (equivalently, from the gluon condensation).

One of the remarkable results from RHIC physics \cite{Braun01} is
the common freezeout temperature of $T_{fo}=174\pm 7$ MeV for all
hadrons. For a $\rho$-meson of mass $m_\rho=770$ MeV this means a
total energy of 1090 MeV, including thermal energy, at 174 MeV. We
neglect the difference between this and the 170 MeV we obtain from
LGS.

At $T_c$, each of the massless quark and antiquark coming together
to make up the $\rho$ will have asymptotic energy $3.15 T$, so the
$\rho$ is formed with energy $6.3 T$. For $T_c=175$ MeV this means
$E_\rho=1103$ MeV, in other words only slightly -- if at all --
more energy than it freezes out at. Thus, the ``dark energy" fed
in from the field energy keeps the temperature essentially
constant.

The critical temperature is the fixed point not only for the
masses but also for the vector coupling $g_V$ which goes to zero
there. Thus hadrons have only weakly interacting, essentially
perturbative coupling until they get most of their masses back;
i.e. go nearly on shell. Thus, they move relatively freely until
nearly back on shell, but kinetic energy is converted into mass.

We believe that the Harada-Yamawaki work shows that the nature of
the chiral symmetry breaking as $T$ goes below $T_c$ is described
well by NJL-type mean field. This does not mean that the
phase-transition is mean field. The fact that the vector meson
masses go to zero is highly suggestive of an $\omega$
condensation~\cite{kurt}; i.e., a density discontinuity in the
transition, so that it is probably first order over most of the
phase diagram.

\section*{Acknowledgments}
We are grateful to Ralf Rapp for useful discussions and for providing
us with data. CHL is supported by Korea Research Foundation
Grant (KRF-2002-070-C00027). GEB and LG were partially supported by the
US Department of Energy under Grant No. DE-FG02-88ER40388

\appendix
\section{Appendix}

The trace (conformal) anomaly is
 \be \theta \equiv T_\mu^\mu = -\langle 0 |
[\beta(g)/2g] (G_{\mu\nu}^a)^2|0\rangle
 \ee
where $\beta(g)=-(g^3/16\pi^2)(11-\frac 23 N_f)$ is the beta
function of QCD at one loop, and $N_f$ is the number of flavors.
The value of the gluon condensate
 \be \frac{g^2}{4\pi^2}\langle 0|
(G_{\mu\nu}^a)^2 |0\rangle = (330\ {\rm MeV})^4 = 0.012\ {\rm
GeV}^4
 \ee is well pinned down from the states of charmonium using
the QCD sum rules \cite{Shifman}. We work in the chiral limit, but
compare with calculations made with both light and heavier
dynamical quarks, so we add the contribution from the quarks which
derives from explicit breaking
 \be \sum m_q\bar q q.
  \ee
The lighter quarks in Miller's LGS of Fig.~\ref{fig1} had the
$MILC$ collaboration's bare mass of $\sim 6$ MeV and the heavier
ones, $\sim 24$ MeV. Since $\langle\bar q q\rangle\sim - (250\
{\rm MeV})^3$,
 \be \frac{m_q\langle\bar q
q\rangle}{\frac{g^2}{4\pi^2} \langle 0| (G_{\mu\nu}^a)^2|0\rangle}
\sim 0.03
 \ee
explaining why no difference is seen from Miller's curve.

It has been somewhat of a mystery why the order parameter of the
chiral restoration transition is $\langle\bar q q\rangle^\star$,
the quark density condensate, whereas the masses of hadrons, e.g.,
the mass of the nucleon is given by the trace anomaly
 \be
m_N=\langle N|\theta|N\rangle.
 \ee
The trace anomaly must be connected by QCD to $\langle\bar q
q\rangle^\star$ in such a way that: {\it In driving a car the
speedometer, i.e. $\langle\bar q q\rangle^\star$, tells one how
fast one is driving, but the car actually moves because the
wheels, i.e. $\theta$, turn.}

In our note above we have described in a simple way the conduit
coupling $\langle\bar q q\rangle$ and $\theta$, in terms of the
condensation energy (i.e., bag constant). Essentially the soft
glue is the glue connected with quarks; the quarks would not have
their masses were the symmetry not broken by the soft glue. Chiral
symmetry restoration implies the melting of the soft glue (the
restoration of scale invariance in the constituent quark sector).
This is precisely what has been found in the skyrmion description
of dense matter in \cite{LPRV}.

Note that the lattice calculation melting the soft glue is carried
out for temperature well within the (effective) hadron sector.
Thus, from the standpoint of the part of the trace anomaly which
vanishes at $T_c$, the order parameter is the hadron mass, in the
case discussed $m_N^\star$, but equally well the $\rho$-meson mass
$m_\rho^\star$ as used by Adami \& Brown\cite{Adami}. The
connection of the hadron mass with $\langle\bar q q\rangle^\star$
is described by the Harada and Yamawaki theory, in that the
scaling of $m_\rho^\star$ towards the fixed point $m_\rho^\star=0$
at chiral restoration is given by $\langle\bar q
q\rangle^\star\rightarrow 0$.
\def\prl{Phys. Rev. Lett}
\def\np{Nucl. Phys.}
\def\pr{Phys. Rev.}

\end{document}